\documentclass{pasj01}
\draft
\usepackage{siunitx, multirow}
\usepackage{graphicx, color}
\usepackage[T1]{fontenc}

\begin{document} 
\Received{}%{yyyy/mm/dd}
\Accepted{}%{yyyy/mm/dd}
%\Published{yyyy/mm/dd}

\title{Development of a new wideband heterodyne receiver system for the Osaka 1.85--m mm--submm telescope\large{\\--- Receiver development \& the first light of simultaneous observations in 230\,GHz and 345\,GHz bands with an SIS--mixer with 4--21\,GHz IF output ---}}

%%% begin:list of authors
% Do NOT capitalize all letters in "textsc".
\author{Sho \textsc{Masui}\altaffilmark{1}}
\author{Yasumasa \textsc{Yamasaki}\altaffilmark{1}}
\author{Hideo \textsc{Ogawa}\altaffilmark{1}}
\author{Hiroshi \textsc{Kondo}\altaffilmark{1}}
\author{Koki \textsc{Yokoyama}\altaffilmark{1}}
\author{Takeru \textsc{Matsumoto}\altaffilmark{1}}
\author{Taisei \textsc{Minami}\altaffilmark{1}}
\author{Masanari \textsc{Okawa}\altaffilmark{1}}
\author{Ryotaro \textsc{Konishi}\altaffilmark{1}}
\author{Sana \textsc{Kawashita}\altaffilmark{1}}
\author{Ayu \textsc{Konishi}\altaffilmark{1}}
\author{Yuka \textsc{Nakao}\altaffilmark{1}}
\author{Shimpei \textsc{Nishimoto}\altaffilmark{1}}
\author{Sho \textsc{Yoneyama}\altaffilmark{1}}
\author{Shota \textsc{Ueda}\altaffilmark{1}}
\author{Yutaka \textsc{Hasegawa}\altaffilmark{1}}
\author{Shinji \textsc{Fujita}\altaffilmark{1}}
\author{Atsushi \textsc{Nishimura}\altaffilmark{1}}
\author{Takafumi \textsc{Kojima}\altaffilmark{2,3}}
\author{Kazunori \textsc{Uemizu}\altaffilmark{2}}
\author{Keiko \textsc{Kaneko}\altaffilmark{2}}
\author{Ryo \textsc{Sakai}\altaffilmark{2}}
\author{Alvaro \textsc{Gonzalez}\altaffilmark{2,3}}
\author{Yoshinori \textsc{Uzawa}\altaffilmark{2,3}}
\author{Toshikazu \textsc{Onishi}\altaffilmark{1}}

%%% end:list of authors

\altaffiltext{1}{Osaka Prefecture University, 1-1 Gakuen-cho, Naka-ku, Sakai, Osaka, 599-8531, Japan}
\email{s\_s.masui@p.s.osakafu-u.ac.jp}
\altaffiltext{2}{National Astronomical Observatory of Japan, 2-21-1 Osawa, Mitaka, Tokyo 181-8588, Japan}
\altaffiltext{3}{Graduate University for Advanced Studies, Mitaka, Tokyo 181-8588, Japan}

 %Do NOT move this preamble from here!

\maketitle

\begin{abstract}

We have developed a wideband receiver system for simultaneous observations in CO lines of {\sl J} = 2--1 and {\sl J} = 3--2 transitions using the Osaka 1.85--m mm--submm telescope.
As a frequency separation system, we developed multiplexers that connect three types of diplexers, each consisting of branch-line couplers and high-pass filters.
The radio frequency (RF) signal is eventually distributed into four frequency bands, each of which is fed to a superconductor-insulator-superconductor (SIS) mixer.
The RF signal from the horn is divided into two frequency bands by a wideband diplexer with a fractional bandwidth of 56\,\%, and then each frequency band is further divided into two bands by each diplexer.
The developed multiplexers were designed, fabricated, and characterized using a vector network analyzer.
The measurement results showed good agreement with the simulation.
The receiver noise temperature was measured by connecting the SIS--mixers, one of which has a wideband 4--21\,GHz intermediate frequency (IF) output.
The receiver noise temperatures were measured to be $\sim$70\,K in the 220\,GHz band, $\sim$100\,K in the 230\,GHz band, 110--175\,K in the 330\,GHz band, and 150--250\,K in the 345\,GHz band.
This receiver system has been installed on the 1.85--m telescope at the Nobeyama Radio Observatory.
We succeeded in the simultaneous observations of six CO isotopologue lines with the transitions of {\sl J} = 2--1 and {\sl J} = 3--2 toward the Orion\,KL as well as the on-the-fly (OTF) mappings toward the Orion\,KL and W\,51.

\end{abstract}

%\newpage

\section{Introduction}
\begin{figure} [h]
     \begin{center}
      \includegraphics[width=10cm]{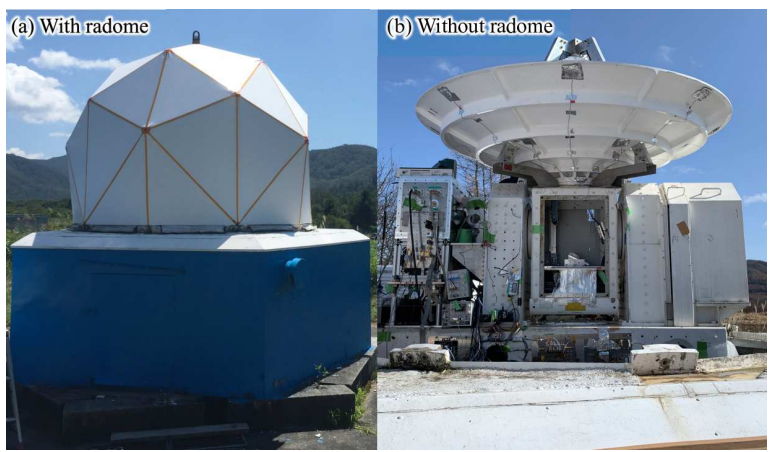} 
     \end{center}
    \caption{(a) The Osaka 1.85-m mm-submm telescope with radome, (b) without radome}\label{photograph_of_1p85m}
    \end{figure}

The 1.85--m mm--submm telescope (figure\,\ref{photograph_of_1p85m}) of Osaka Prefecture University (OPU) has been operated at the Nobeyama Radio Observatory (NRO) to observe molecular clouds in nearby star-forming regions and along the Galactic Plane in the molecular lines of ${}^\text{12}$CO, ${}^\text{13}$CO, and C${}^\text{18}$O ({\sl J} = 2--1) (\cite{Onishi_1p85m}, \cite{Nishimura_1p85m}, \yearcite{Nishimura_SPIE}).
A receiver system based on superconductor--insulator--superconductor (SIS) mixers has been installed in the telescope in order to achieve highly sensitive and efficient observations with low noise temperature close to the several times quantum limit.
Now, we are planning to relocate the 1.85-m telescope to a site (altitude $\sim$2,500\,m) at the Atacama Desert in Chile and to newly install a wideband receiver for simultaneous observations of CO isotopologue lines with the transitions of {\sl J} = 2--1, {\sl J} = 3--2 and other molecular lines.
Historically, the quasi-optical system has been used to achieve the simultaneous observations in such widely separated molecular lines in frequency (e.g., \cite{Kosma_3m}).
This is because the large fractional bandwidth and wideband intermediate frequency (IF) system had been difficult to be developed only with the waveguide system.  
As a frequency separation system, several types of multiplexers have been proposed and developed so far (\cite{Han_diplexer}, \cite{Okada_hinotori}, \cite{Kojima_multiplexer}, \cite{Nakajima_multiplexer}).
\citet{Han_diplexer} and \citet{Okada_hinotori} have developed receivers using quasi-optical diplexers.
The quasi-optical diplexer enables simultaneous observations of molecular spectra with widely separated frequencies.
However, the optics becomes large and complicated, which makes the system difficult to implement inside the front-end cryostat or even into the telescope optical path. Therefore, quasi-optical diplexers are usually installed at room temperature, with the consequent increase in noise temperature due to extra loss in the insertion path.
\citet{Kojima_multiplexer} and \citet{Nakajima_multiplexer} have been developed waveguide multiplexers composed of several diplexers.
These multiplexers made the receiver system compact and simple since the system can be realized with a single horn.
Therefore, we decided to develop a waveguide multiplexer that separates 230\,GHz and 345\,GHz bands, a wideband optical system, and a corrugated horn for 210--375\,GHz band.
The development of the wideband optics with the corrugated horn is described in a separate paper (Yamasaki et al. 2021).

In recent years, the National Astronomical Observatory of Japan (NAOJ) developed a wideband SIS-mixer with 4--21\,GHz IF output (\cite{Kojima_wideband_RF_IF}).
This wideband IF system will enhance the flexibility of the IF system and thus the ability to observe many scientifically desired molecular lines simultaneously.
For example, the ${}^\text{12}$CO, ${}^\text{13}$CO, and C${}^\text{18}$O molecular lines in 345\,GHz band can be observed simultaneously in one side-band of the SIS--mixer with the 4--21\,GHz IF output.
Another advantage of the wide IF mixer is that the LO only needs to cover a narrower frequency range for the same radio frequency (RF) coverage.
We decided to use this SIS--mixer to achieve our aimed observations.

In this paper, we report the development of the wideband receiver system using the waveguide multiplexers, including the wideband SIS--mixer with IF output of 4--21\,GHz, and the measurement results of the system.
In addition, we also report the results of the first light observations with the 1.85--m telescope using the new wideband receiver system.

\section{Development of a wideband waveguide multiplexer system}

\subsection{Introduction to the waveguide multiplexer system}

There are several types of waveguide multiplexers, such as hybrid-coupled, manifold-coupled, circulator-coupled, and directional filter multiplexers.
\citet{Cameron_manifold} described a comparison of these multiplexers.
Manifold-coupled, circulator-coupled, and directional filter types (\cite{Ruiz-Cruz_triplexer}, \cite{directional_filter}) are compact and have several advantages than the hybrid-coupled type.
However, these multiplexers have a complex structure, narrow bandwidth, and high insertion loss.

The hybrid-coupled multiplexer can be made by cascade connections of diplexers (\cite{Asayama_diplexer}, \cite{Hasegawa_diplexer}) that consist of branch-line coupler (BLC) and band-pass filter (BPF).
These multiplexers were designed to separate a fractional bandwidth of about 30\,\% (between minimum frequency and maximum frequency) in the RF band into 3 or 4 bands (\cite{Kojima_multiplexer}, \cite{Nakajima_multiplexer}) to achieve the simultaneous observations of molecular lines.
In their multiplexers, the signals are extracted in the order of high frequency to low frequency.
This means that lower frequency signals pass through more diplexers, indicating that the lower the frequency band is, the longer the path length and thus the greater the insertion loss is.
Since the waveguide insertion loss is lower at lower frequencies, this compensates to some extent the longer path length, obtaining flatter insertion loss within the RF band.

\subsection{Conceptual design of the wideband waveguide multiplexer}

\begin{figure} [h]
     \begin{center}
      \includegraphics[width=15cm]{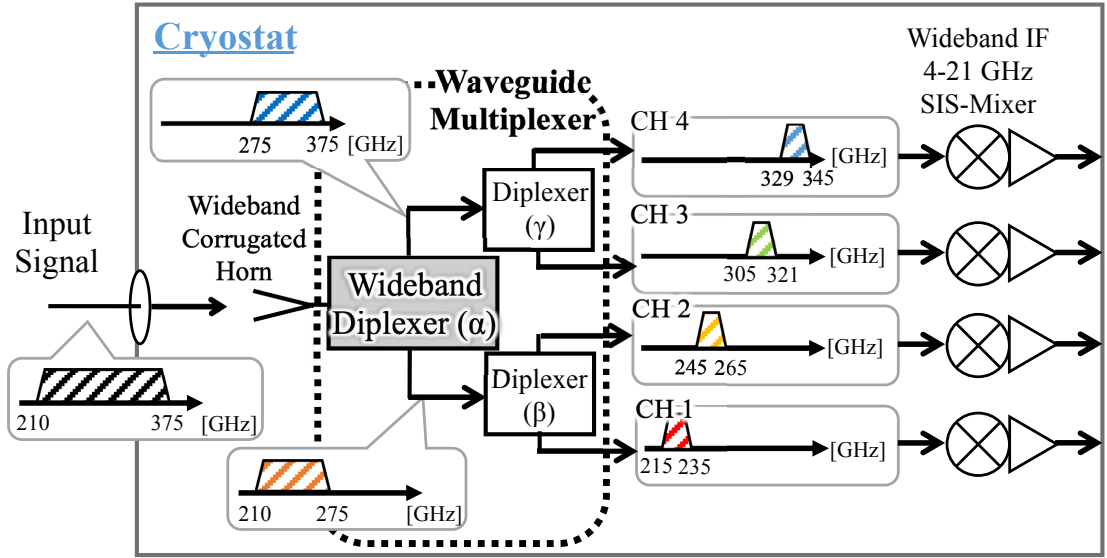} 
     \end{center}
    \caption{Schematic diagram of the final configuration of the planned new receiver system for the simultaneous observations in CO isotopologue lines of the transitions of {\sl J} = 2--1 and {\sl J} = 3--2. The receiver is mainly composed of three components;  a wideband corrugated horn, a waveguide multiplexer, and wideband IF SIS-mixers. The waveguide multiplexer consists of a wideband diplexer\,($\alpha$) and two diplexers\,($\beta$), ($\gamma$). The wideband diplexer\,($\alpha$) provides frequency separation between 210--275\,GHz and 275--375\,GHz. The two output ports are connected to each diplexer\,($\beta$) and ($\gamma$), respectively, and eventually separated into four frequency bands; 215--235\,/\,245--265\,/\,305--321\,/\,329--345\,GHz.
    }\label{fig:Schematic_Diagram}
    \end{figure}

Here, we show the conceptual design of the waveguide multiplexer with a fractional bandwidth of more than 50\%.
As shown in figure\,\ref{fig:Schematic_Diagram}, the wideband diplexer splits the RF signal from the horn into two frequency bands with a wide bandwidth.
Then, each frequency band is divided into two bands by each diplexer.
This makes it possible to unify the number of diplexers to be passed through to two for all channels in contrast to the cascade connection type.
The multiplexers consist of the wideband diplexer\,($\alpha$) to separate the two RF bands, 210--275\,\si{\giga\hertz} and 275--375\,\si{\giga\hertz}, and the following two diplexer\,($\beta$) and diplexer\,($\gamma$) focused on lines of CO isotopologues with the transitions of {\sl J} = 2--1 and {\sl J} = 3--2, respectively, to achieve dual side-band observations for each RF band.
The wideband SIS--mixers with 4--21\,GHz IF output (\cite{Kojima_wideband_RF_IF}) are expected to be connected to 4 output channels of the wideband multiplexer and driven by two LO sources at fixed frequencies of 240\,GHz and 325\,GHz.

It has been quite challenging to develop the wideband diplexer\,($\alpha$) that can separate the fractional bandwidth over 50\,\%.
In \citet{Gonzalez_diplexer}, a prototype wideband diplexer with combining LO signal was developed and fabricated in 275--500\,GHz band.
Referring to the model of \citet{Gonzalez_diplexer}, we optimized our system as an RF circuit of 210--375\,GHz band.
The design of the wideband diplexer, including wideband BLC and high-pass filter (HPF), is described in detail in \citet{Masui_diplexer}.

Each output port of the diplexer\,($\beta$) and diplexer\,($\gamma$) is connected to the IF 4--21\,GHz SIS--mixer.
Therefore, it is necessary to design the diplexers with a passband of 21\,GHz.
In addition, the passband of the diplexers are set to be larger than 21\,GHz in order to accommodate further expansion of the IF band of SIS--mixer.
To achieve this condition, we applied a HPF type diplexer (\cite{Hasegawa_diplexer}), not a BPF type diplexer used in conventional multiplexers.

\subsection{Design and specifications of the waveguide multiplexer}
\begin{figure} [h]
     \begin{center}
      \includegraphics[width=15cm]{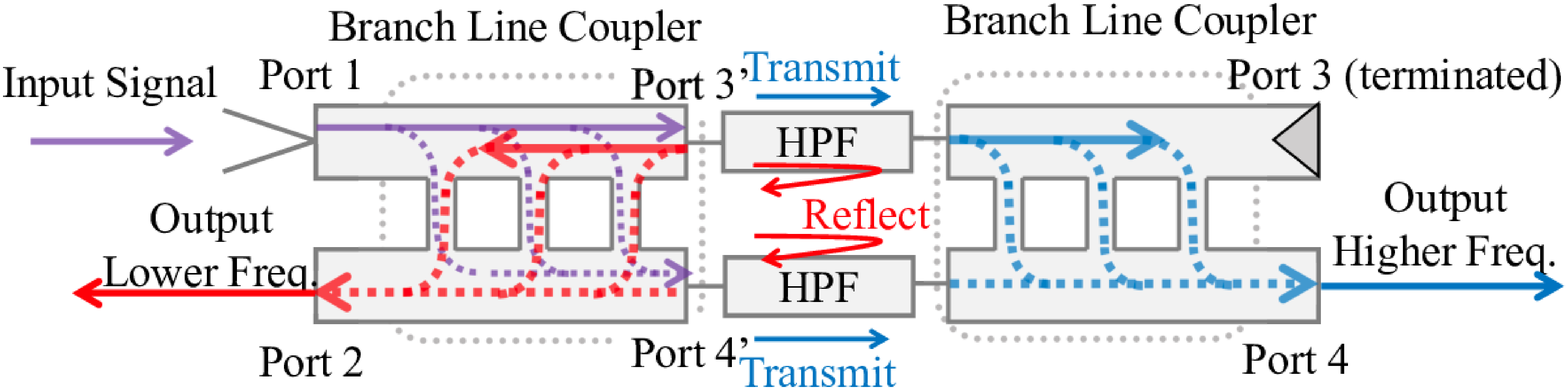} 
     \end{center}
    \caption{Schematic diagram of the diplexer we developed. It is composed of two branch-line couplers (BLCs) and two high-pass filters (HPFs).
    }\label{fig:schematic_diagram_of_diplexer}
    \end{figure}

Figure\,\ref{fig:schematic_diagram_of_diplexer} shows a schematic diagram of an HPF--type diplexer, consisting of two BLCs and two HPFs.
The BLCs divide the input radio waves into the same intensity of $-$3\,dB with a phase difference of 90 degrees, while the HPFs pass the radio waves higher than the cutoff frequency and reflect the lower ones.
Therefore, when radio wave is input to the diplexer from port\,1, the radio wave higher than the cutoff frequency is output from port\,4, and the radio wave lower than the cutoff frequency is output from port\,2.
Port\,3 is terminated by a 4\,K load that absorbs unwanted RF signal and LO power. The load contributes to the receiver noise temperature on ports 2 and 4 as a source of RF thermal noise.  Since the temperature is cooled to 4\,K,  the load has little effect on the noise temperature; the thermal noise is very weak compared to the intensity of the incoming radio signal from port\,1.
%Port\,3 is terminated.
In order to design three diplexers that make up the wideband multiplexer, the BLC and HPF of each are simulated and optimized.

\begin{figure} [h]
     \begin{center}
      \includegraphics[width=13cm]{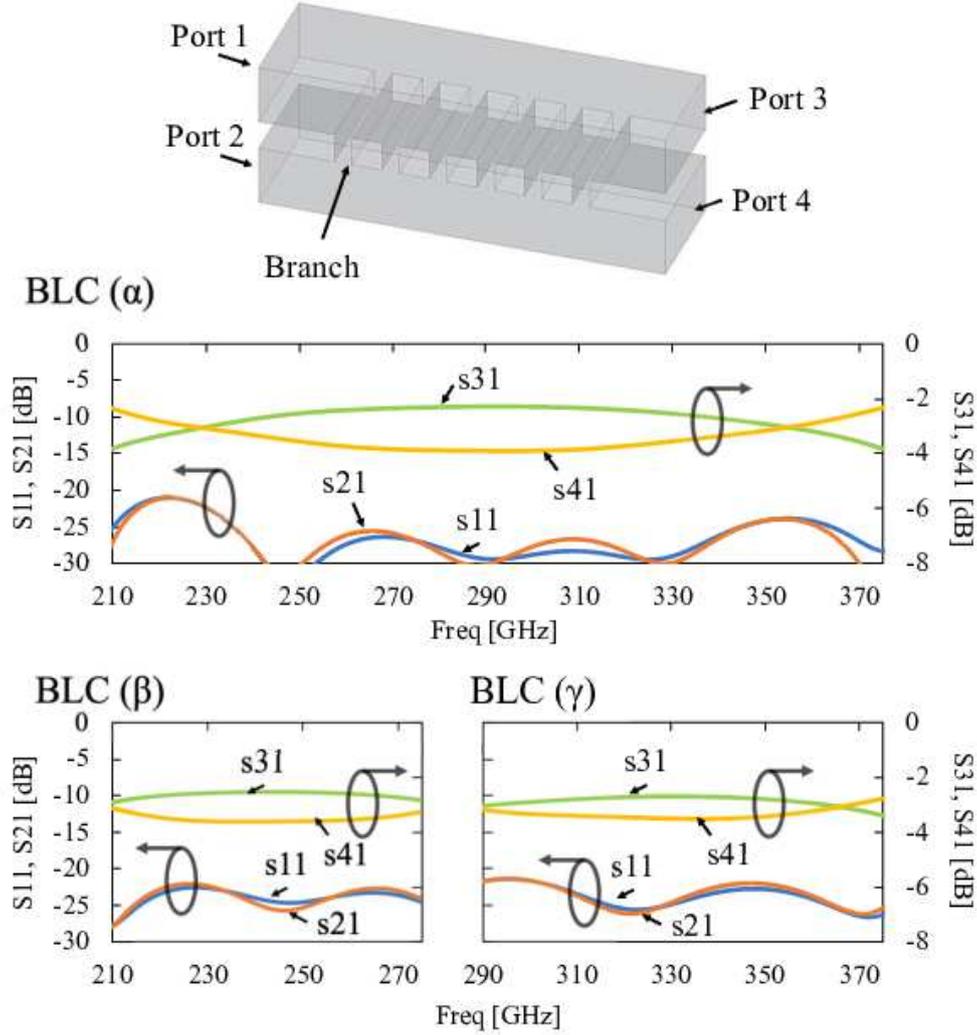} 
     \end{center}
    \caption{
    (Top) Conceptual structure of branch-line coupler (BLC). The number and width of branches vary from circuit to circuit. ($\alpha$) The simulation results of the wideband BLC, ($\beta$) 210--275\,GHz band BLC, and ($\gamma$) 290--375\,GHz band BLC. The sizes of the BLCs are shown in detail in the Appendix.
    }\label{fig:simulation_BLC}
    \end{figure}

First, the design of the BLC is explained.
As mentioned above, in a BLC, it is important to divide the input intensity in half, but it is not possible to make both S31 and S41 precisely equal to $-$3.0\,dB at all frequencies.
In this development, we designed the wideband diplexer\,($\alpha$) BLC with S31 and S41 to be $-$3.0\,$\pm$\,0.75\,dB, and the diplexer\,($\beta$) and ($\gamma$) BLC with S31 and S41 to be $-$3.0\,$\pm$\,0.5\,dB.
Furthermore, the return losses S11 and S21 of each BLC were set to be better than 20\,dB.
The target characteristics of each BLC are summarized in table\,\ref{tab:goal value of BLCs}.
We optimized the design of the diplexers by analyzing the {\sl S}--parameters using the electromagnetic field analysis software HFSS.
Figure\,\ref{fig:simulation_BLC} shows the simulation results of these BLCs.
In the optimized design of the wideband BLC\,($\alpha$), the S31 and S41 is simulated to be $-$3.0\,$\pm$\,0.81\,dB, and the S11 and S21 better than $\sim$21\,dB; for BLC ($\beta$) and ($\gamma$), the S31 and S41 $-$3.0\,$\pm$\,0.54\,dB and $-$3.0\,$\pm$\,0.41dB, respectively, and the S11 and S21 better than $\sim$22 dB.

\begin{table}[h]
  \tbl{The goal characteristics of the BLCs.}{%
  \begin{tabular}{llll}
      \hline
      Component & Parameter &  Goal\\ 
      \hline
       \multirow{2}{*}{Wideband BLC ($\alpha$)}& $\Delta$\,=\,S31$-$S41 & 1.5\,dB \\
       & S11, S21 & $\sim$20\,dB\\ 
       \multirow{2}{*}{200\,GHz BLC ($\beta$)} & $\Delta$\,=\,S31$-$S41 & 1.0\,dB \\ 
       & S11, S21 & $\sim$20\,dB\\ 
       \multirow{2}{*}{300\,GHz BLC ($\gamma$)}& $\Delta$\,=\,S31$-$S41 & 1.0\,dB \\
       & S11, S21 & $\sim$20\,dB\\ 
      \hline
    \end{tabular}}\label{tab:goal value of BLCs}
\begin{tabnote}
%The target value in the design of the characteristic required for each circuit is shown.
\end{tabnote}
\end{table}

Second, we discuss the design of three types of HPFs\,($\alpha$), ($\beta$), and ($\gamma$).
To design the wideband HPF\,($\alpha$), a compact cavity filter (\cite{Rosenberg_pseudo_HPF}) was applied to achieve a large passband, sharper cutoff, and low insertion loss.
The cutoff frequency was set to $\sim$275\,GHz, and the return loss was set to be better than 20\,dB.
The cutoff frequencies of HPFs\,($\beta$) and ($\gamma$) were set to $\sim$238\,GHz and $\sim$323.5\,GHz, respectively, and the return loss was set to be better than 20\,dB.
The target characteristics of each HPF are summarized in table\,\ref{tab:goal value of HPFs}.
The models of the HPFs\,($\alpha$), ($\beta$) and ($\gamma$) are shown in figure\,\ref{fig:simulation_HPF}, and the simulation results are presented below.
In the final design of the wideband HPF\,($\alpha$), the cutoff frequency is about 277\,GHz, and the return loss is better than $\sim$21\,dB; for HPFs\,($\beta$) and ($\gamma$), the return losses are better than $\sim$20\,dB in most of the bands.
In some frequency range of HPF\,($\beta$), the return loss was $\sim$18\,dB.

\begin{table}[t]
  \tbl{The goal characteristics of the HPFs.}{%
  \begin{tabular}{llll}
      \hline
      Component & Parameter &  Goal\\ 
      \hline
       \multirow{2}{*}{Wideband HPF ($\alpha$)}& Return loss & $\sim$20\,dB \\
       & cutoff frequency & $\sim$275\,GHz\\
       \multirow{2}{*}{200\,GHz HPF ($\beta$)}& Return loss & $\sim$20\,dB \\
       & cutoff frequency & $\sim$238\,GHz\\
       \multirow{2}{*}{300\,GHz HPF ($\gamma$)}& Return loss & $\sim$20\,dB \\
       & cutoff frequency & $\sim$323.5\,GHz\\
      \hline
    \end{tabular}}\label{tab:goal value of HPFs}
\begin{tabnote}
%The target value in the design of the characteristic required for each circuit is shown.
\end{tabnote}
\end{table}

    \begin{figure} [h]
     \begin{center}
      \includegraphics[width=10cm]{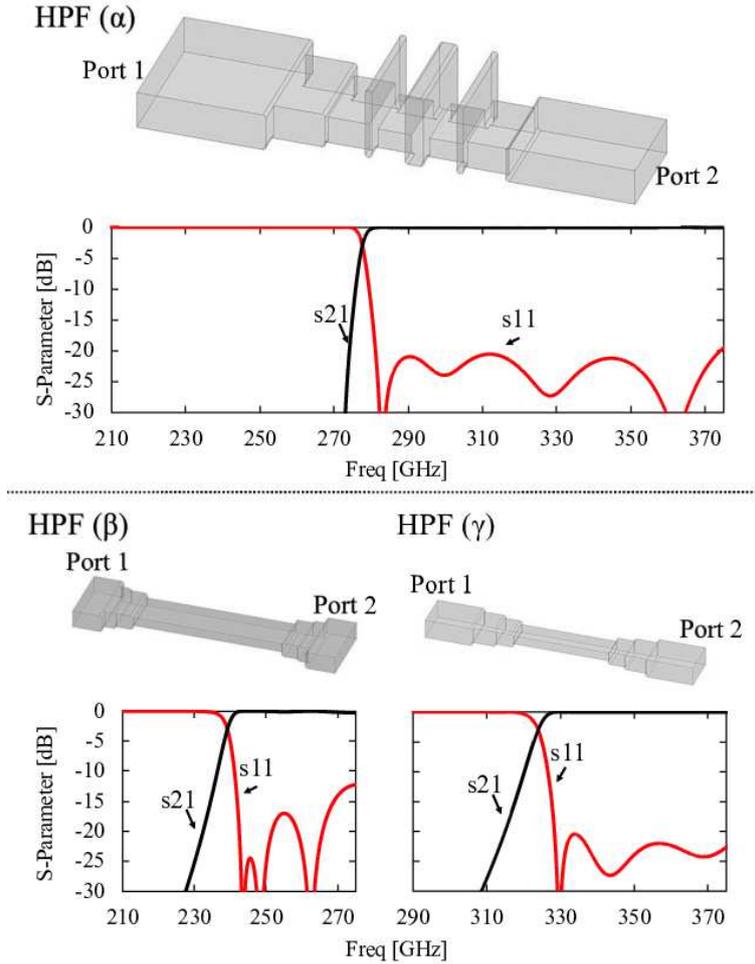} 
     \end{center}
    \caption{
    The model and simulation results of the designed HPFs. ($\alpha$) the HPF of the wideband diplexer ($\alpha$), ($\beta$) the 210--275 GHz band HPF, and ($\gamma$) the 290--375 GHz HPF. The sizes of the HPFs are shown in detail in the Appendix.}
    \label{fig:simulation_HPF}
    \end{figure}

\begin{figure*} [t]
 \begin{center}
  \includegraphics[width=16cm]{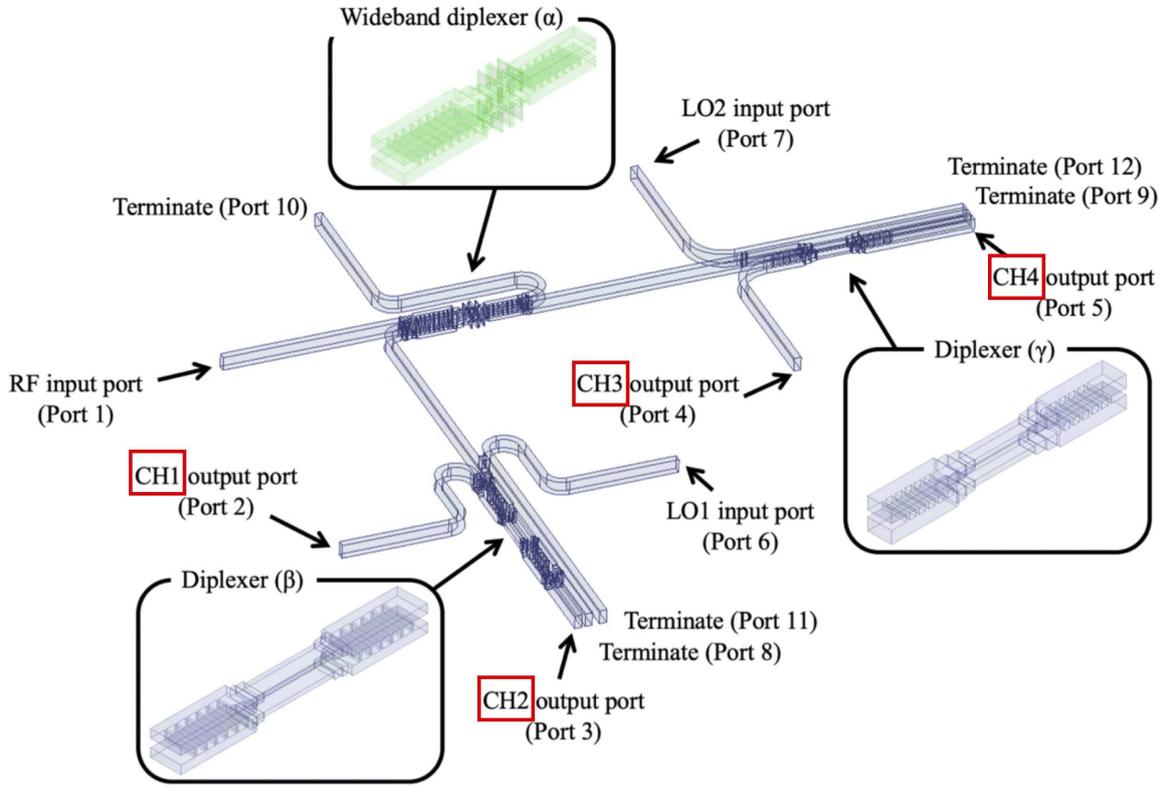} 
 \end{center}
\caption{
The layout of the multiplexer. It consists of a wideband 210--375\,GHz diplexer\,($\alpha$), a 210--275\,GHz diplexer\,($\beta$), and a 290--375\,GHz diplexer\,($\gamma$). This shows how three diplexers are connected. Enlarged views of the diplexers are also shown. A waveguide circuit that supplies LO is connected to each of the two diplexers\,($\beta$), ($\gamma$), which is then mixed with the RF signal and fed to the connected SIS--mixers. 
}\label{fig:SimulationCAD}
\end{figure*}

\begin{figure} [h]
 \begin{center}
  \includegraphics[width=15cm]{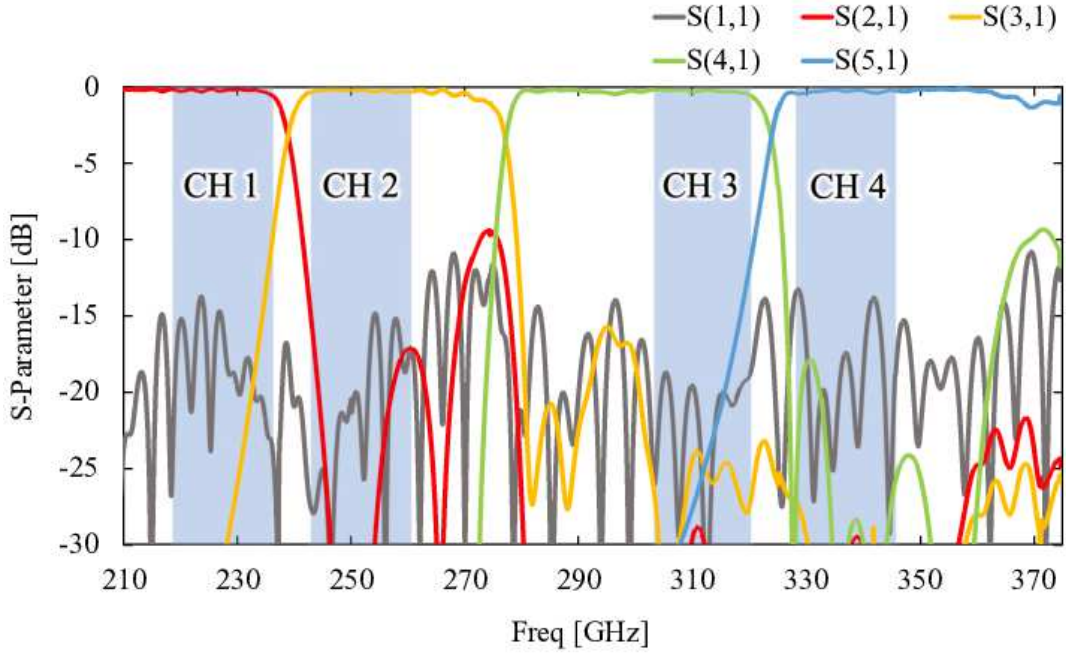} 
 \end{center}
\caption{
The simulation results of the designed multiplexer. The conductivity of the PEC is assumed in the simulation. The blue square area indicates the bandwidths to be used in the observations.
}\label{fig:Simulation_result_of_multiplexer}
\end{figure}

The configuration of the multiplexer was made to connect each designed BLC to the HPF as shown in figure\,\ref{fig:SimulationCAD}.
In order to connect the developed multiplexers to the SIS--mixers and to mix them with LO in the SIS--mixers, LO couplers were built into the input ports of the diplexers\,($\beta$) and ($\gamma$).
LO couplers have a coupling ratio of about 17\,dB and a directivity of more than 20\,dB in the frequency range of interest.
The input LO signals are transmitted to each output port of the diplexers\,($\beta$) and ($\gamma$) in the same way as RF signals.
The size of the input\,/\,output waveguide is set to WR--3.15 (0.8\,$\times$\,0.4\,\si{\milli\metre}) in order to prevent higher-order modes of the waveguide from occurring; the cutoff frequencies of the fundamental and the first higher-order modes are 187.5\,GHz and 375\,GHz, respectively.
Note that the "WR" designation stands for Rectangular Waveguides, and the number that follows "WR" is the width of the waveguide opening in mils, divided by 10 (\cite{Hesler_WR}).
To connect these components, impedance matching circuits are connected.
The simulation results are shown in figure\,\ref{fig:Simulation_result_of_multiplexer}, where the four frequency bands corresponding to IF 4--21\,GHz are shown with a blue square area.
This analysis is calculated using the conductivity of a perfect electrical conductor (PEC).
The total return loss is simulated to be $\sim$15\,dB.
We also simulated the image rejection ratios, and they are designed to be better than $\sim$20\,dB in most of the bands.
The sizes of the diplexers are shown in detail in the Appendix.

\subsection{Fabrication of waveguide multiplexer}
\begin{figure} [h]
 \begin{center}
  \includegraphics[width=15cm]{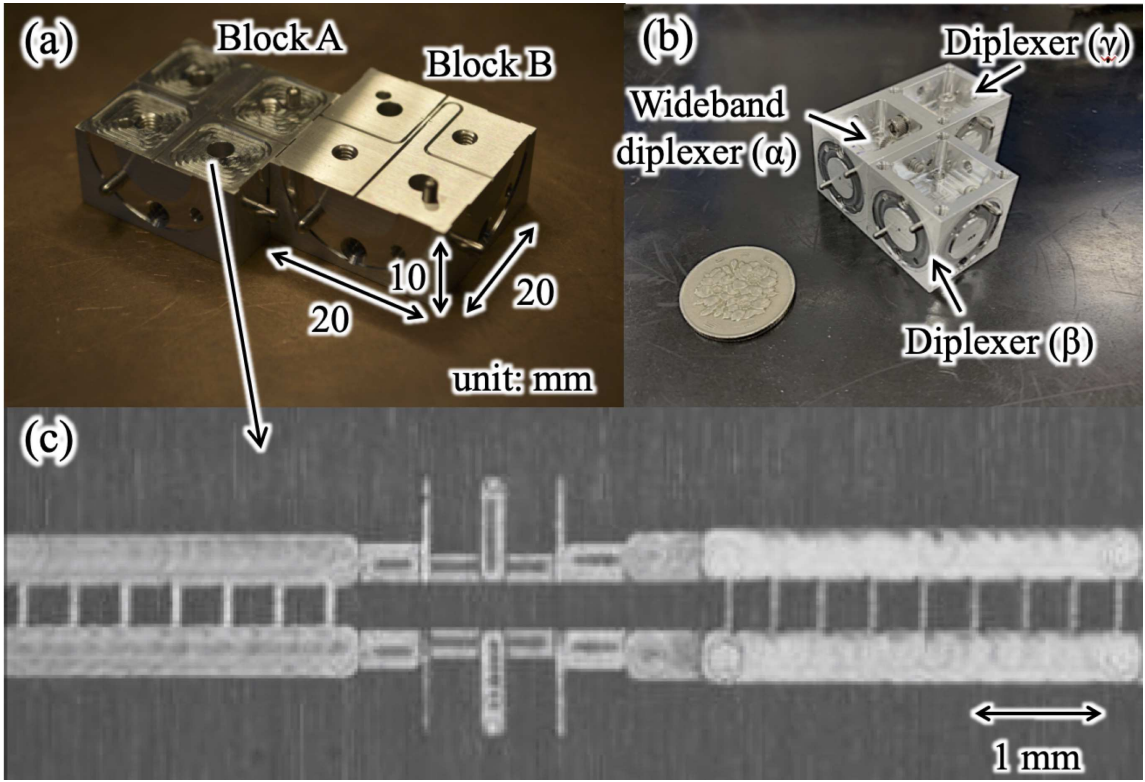} 
 \end{center}
\caption{
(a) Photograph of the split-blocks, described as Block A and Block B, of the fabricated wideband diplexer($\alpha$). (b) Photograph of the assembled wideband multiplexer taken with a 100 Japanese yen coin. (c) Detailed optical image of the circuit part.
}\label{fig:Block_of_multiplexer}
\end{figure}

The wideband diplexer\,($\alpha$), and diplexers\,($\beta$) and ($\gamma$) were fabricated in separate blocks to accommodate future changes as the observed frequency.
Each block is composed of two E--plane split--blocks, and the waveguide circuits were fabricated by direct machining using an end mill.
Aluminum alloy A6063, which is easy to machine and has excellent electrical conductivity, was used for these split--blocks.
In particular, the conductivity of A6063 at 4\,K is $\sim4.0\times{10}^\text{8}$\,S/m (\cite{Aluminum_alloy_property}), which reduces the insertion loss of the waveguide circuit significantly at low temperature.
An anti-cocking type waveguide flange was used.
The size of the wideband diplexer\,($\alpha$) and diplexer\,($\beta$) blocks was $20\times20\times20$\,\si{\milli\metre} when the two split--blocks were assembled.
The size of the diplexer\,($\gamma$) block was set to $20\times22\times20$\,\si{\milli\metre} when assembled.
Figure\,\ref{fig:Block_of_multiplexer}\,(a) shows a photograph of the manufactured diplexer\,($\alpha$).
Figure\,\ref{fig:Block_of_multiplexer}\,(c) shows an optical image focused on the circuit of the diplexer\,($\alpha$) taken with a Zygo NewView 8000, and figure\,\ref{fig:Block_of_multiplexer}\,(b) shows the waveguide multiplexer with these diplexers assembled.

\subsection{Measurement system of waveguide multiplexer}

\begin{figure} [h]
 \begin{center}
  \includegraphics[width=15cm]{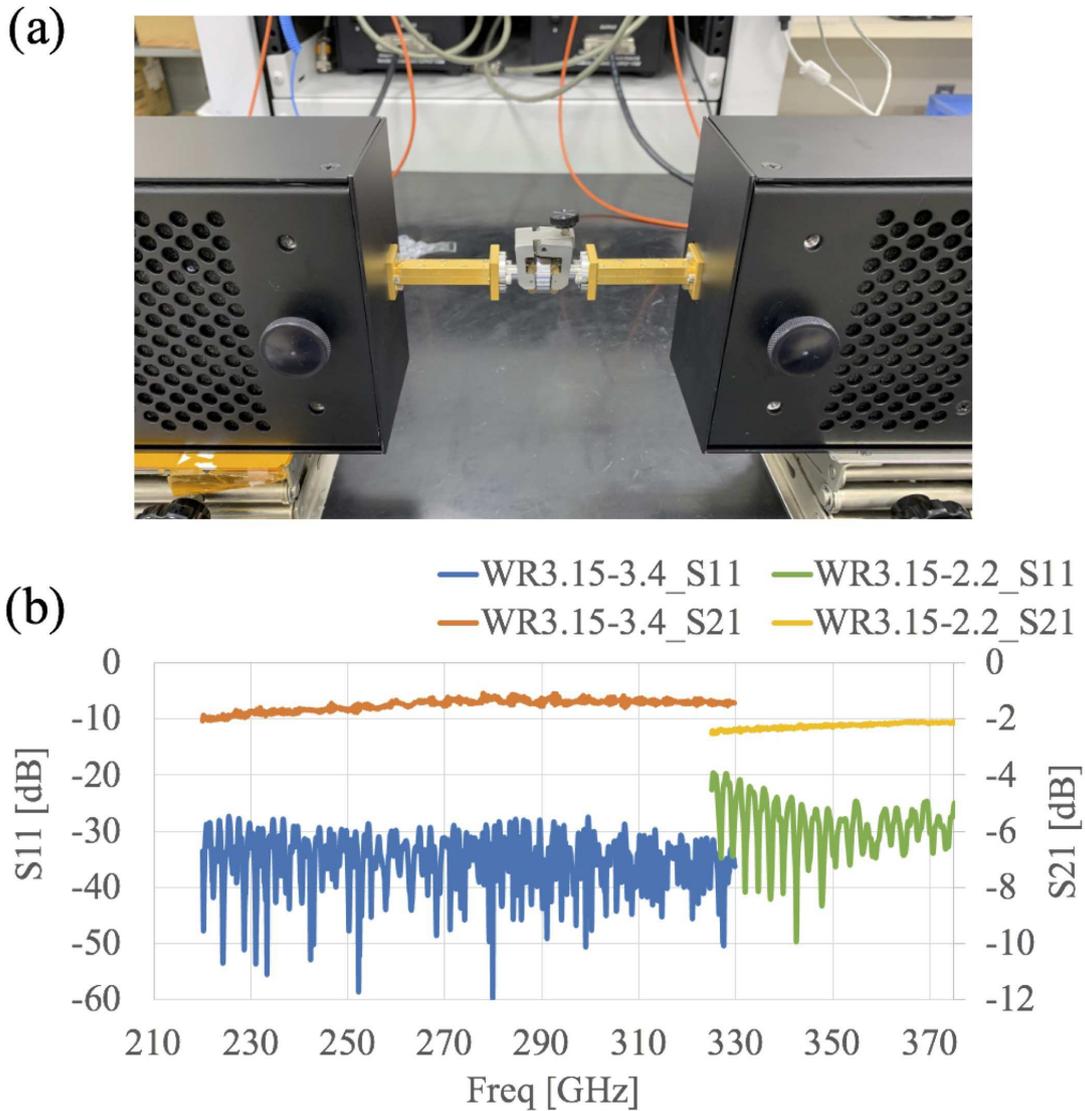} 
 \end{center}
\caption{
(a) Photograph of the setup to measure {\sl S}--parameters for the WR--3.15 to WR--2.2 waveguide transition in the back-to-back configuration using WR--2.2 VNA extender. (b) Measurement results of the WR--3.15 to WR--3.4 transition and the WR--3.15 to WR--2.2 transition using WR--3.4 and WR--2.2 extenders, respectively.
}\label{fig:Transition_meas}
\end{figure}

A PNA-X vector network analyzer (VNA) and waveguide extenders, VNA extension modules, from the National Institute of Information and Communications Technology (NICT) were used for the multiplexer and waveguide transitions measurements.
To ensure a wide frequency range of 210--375\,GHz, a WR--3.4 (0.864\,$\times$\,0.432\,\si{\milli\metre}) extender was used for 220--330\,GHz, and a WR--2.2 (0.559\,$\times$\,0.279\,\si{\milli\metre}) extender was used for 325--500\,GHz.
In order to minimize the mismatch at the flange interface with the extenders, tapered transitions with waveguide sizes from WR--3.15 to WR--3.4 and from WR--3.15 to WR--2.2 with a length of 20\,mm were prepared.
The relatively long waveguide was chosen for this measurement because of its good return loss characteristics.
A picture of a typical measurement setup for back-to-back measurements of the {\sl S}--parameters of the WR--3.15 to WR--2.2 transition using the WR--2.2 extender is shown in figure\,\ref{fig:Transition_meas}\,(a).
Prior to the measurements, a short-open-load-through (SOLT) calibration, a standard calibration method of the VNA and extenders, of two ports was performed.
The measurement results of the waveguide transitions with each extender are shown in figure\,\ref{fig:Transition_meas}\,(b): the return loss and insertion loss for the WR--3.15 to WR--3.4 transition are 28\,dB and 2\,dB, respectively, and those for the WR--3.15 to WR--2.2 transition are better than 20\,dB and 2.5\,dB, respectively.

\subsection{Measurement results of the waveguide multiplexer}

\begin{figure} [p]
 \begin{center}
  \includegraphics[width=15cm]{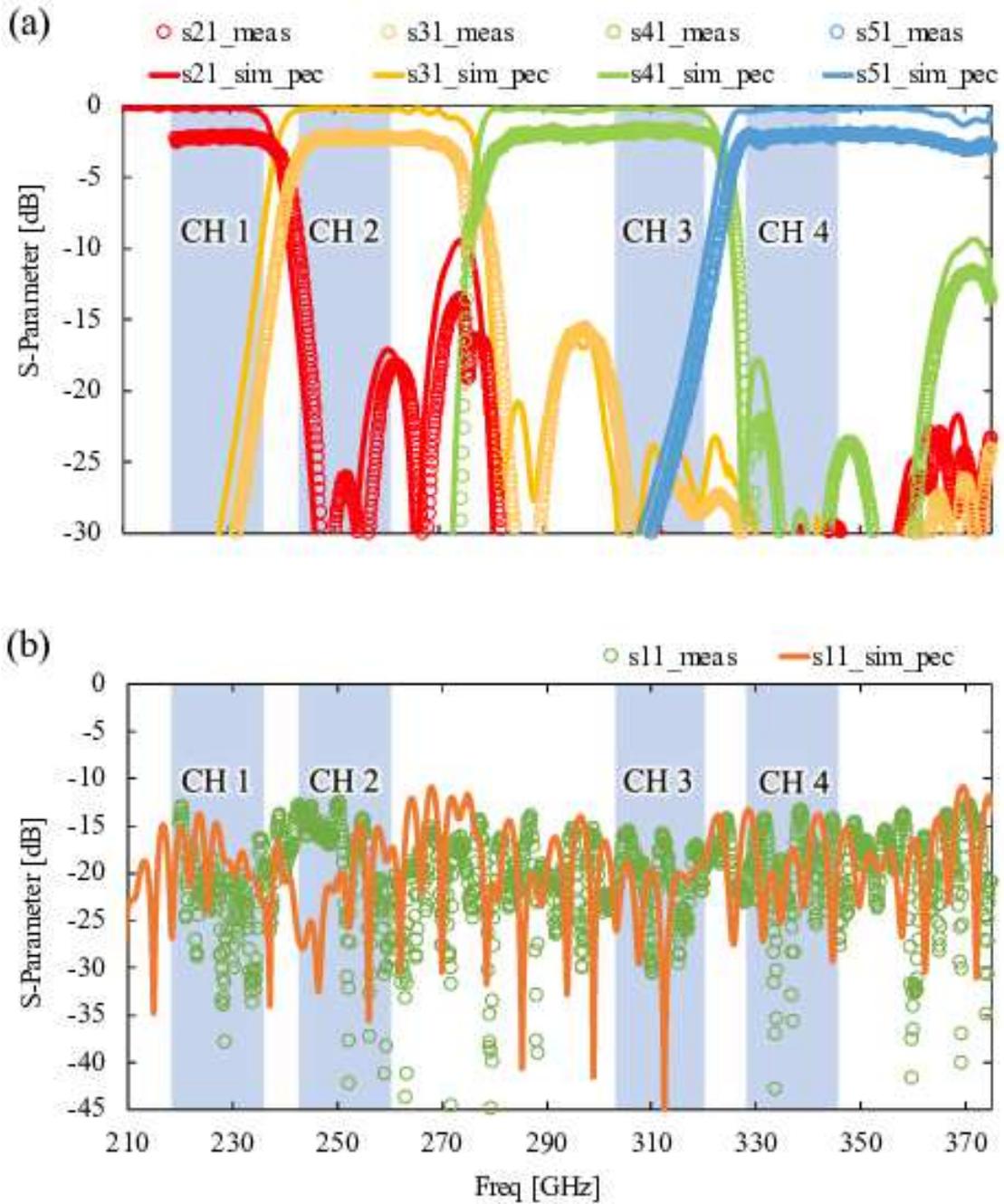} 
 \end{center}
\caption{
Comparison of the simulation and the measurement of the {\sl S}--parameters of the multiplexer.  The measurement was carried out using VNA, and the results are indicated in open circles.  The simulation assumes the conductivity of PEC, and the results are in solid lines. (a) The output levels to Ports 2, 3, 4, and 5 are expressed as S21, S31, S41, and S51, respectively. (b) The return loss, S11. The simulation (solid line) and measurement results (open circle) show good agreement, and the return loss is measured to be better than about 15\,dB throughout the required frequency bands.
}\label{fig:Measurement_result_of_mupltiplexer3}
\end{figure}

The measurement results of the multiplexer are shown in figure\,\ref{fig:Measurement_result_of_mupltiplexer3}.
For S21, S31, S41, and S51, the insertion loss of the waveguide transition described in the previous section are subtracted.
The simulation results using the conductivity of PEC are also plotted as solid lines.
The cutoff frequency of the diplexer\,($\beta$) in the 210--275\,GHz band is shifted to a higher frequency of about 3\,GHz.
We performed the 3-D measurement of the waveguide and found that the waveguide size became smaller, which caused the frequency shift of 3\,GHz.
The return loss was measured to be $\sim$15 dB, which is consistent with the simulation results.

\begin{figure} [h]
 \begin{center}
  \includegraphics[width=15cm]{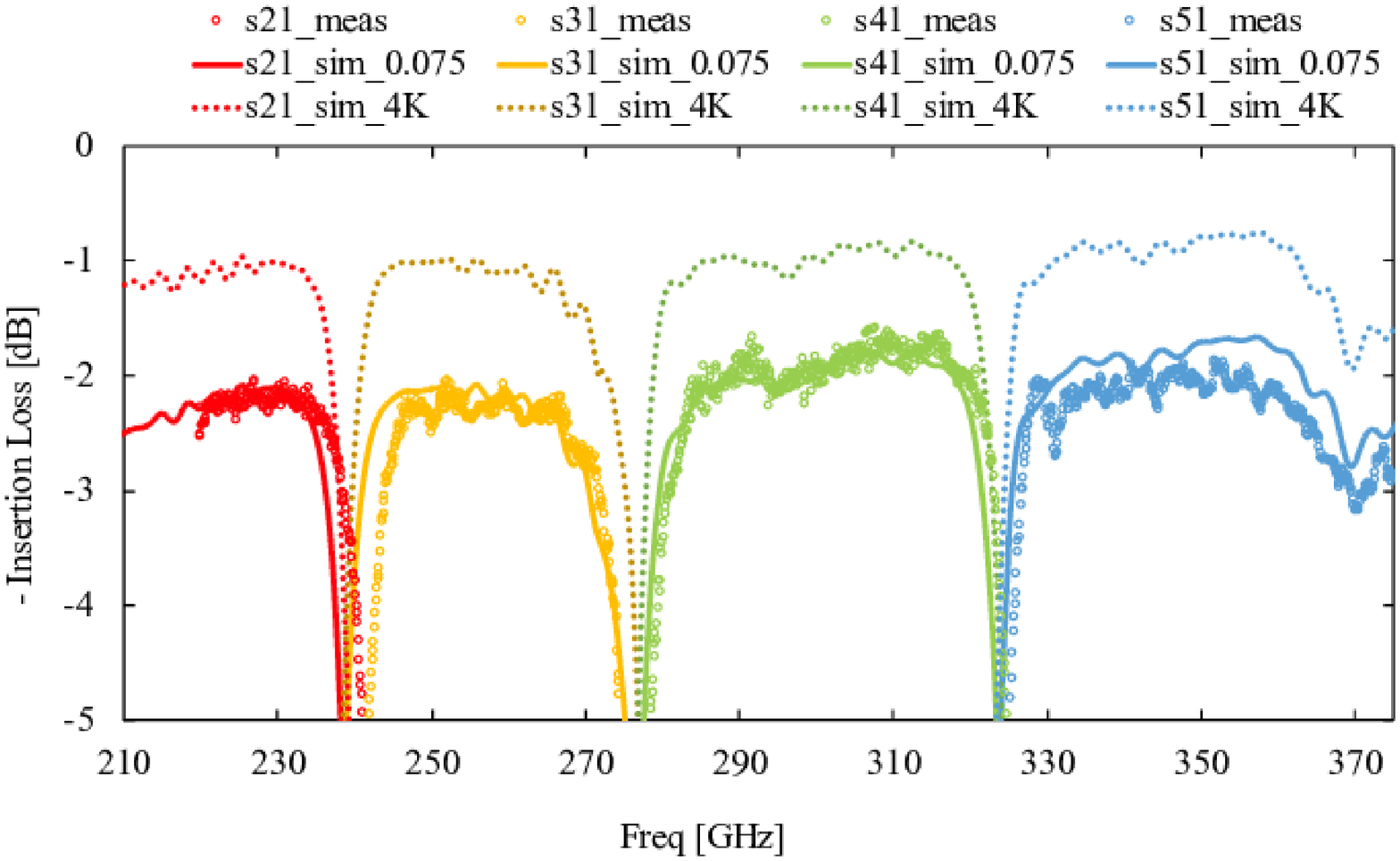} 
 \end{center}
\caption{
A detailed view of the measured S21, S31, S41, and S51 with a simulation assuming a room temperature/4\,K conductivity and the realistic roughness of the waveguide surface. The open circles are the measurement results. The solid lines are the results from the simulation assuming the 300\,K conductivity and the 0.075\,\si{\micro\metre} surface roughness. The measurement results with the open circles, measured at 300\,K, are found to well delineate the solid lines. The dotted lines are the simulation results with the conductivity at 4\,K of aluminum and the roughness of the waveguide surface 0.075\,\si{\micro\metre}. This indicates that the {\sl S}--parameters of the manufactured multiplexer at 4\,K are about 1\,dB over the necessary frequency band.
}\label{fig:Measurement_result_of_mupltiplexer2}
\end{figure}

Here, we discuss the insertion loss by also taking the surface roughness into account.
We assume the surface roughness of the waveguide to be 0.075\,\si{\micro\metre} from \citet{Kojima_multiplexer} and aluminum conductivity of $3.8\times{10}^\text{7}$\,S/m.
The aluminum A6063 is known to exhibit good conductivity at low temperatures at 4\,K (\cite{Aluminum_alloy_property}).
With the above-assumed surface roughness, we simulated the insertion loss using the HFSS both at room temperature and 4K, and plot the results in figure\,\ref{fig:Measurement_result_of_mupltiplexer2} with the measurement results.
This figure shows that the simulation with the room temperature well reproduces the measurement results, indicating that the surface roughness of the waveguide is estimated to be $\sim$0.075\,\si{\micro\metre}.
At 4\,K with the surface roughness, the simulation results show the insertion loss of the multiplexer is estimated to be $\sim$1\,dB.
For example, if the SIS--mixer is used with an SSB noise temperature of about 60\,K, the effect of 1\,dB can be calculated to be $\sim$15\,K, indicating that the effect of insertion loss of the multiplexer is small, though not negligible.

\section{Application to the 1.85m telescope: simultaneous observations of CO in 230\,GHz and 345\,GHz bands}
\subsection{Receiver configuration}

\begin{figure} [h]
 \begin{center}
  \includegraphics[width=15cm]{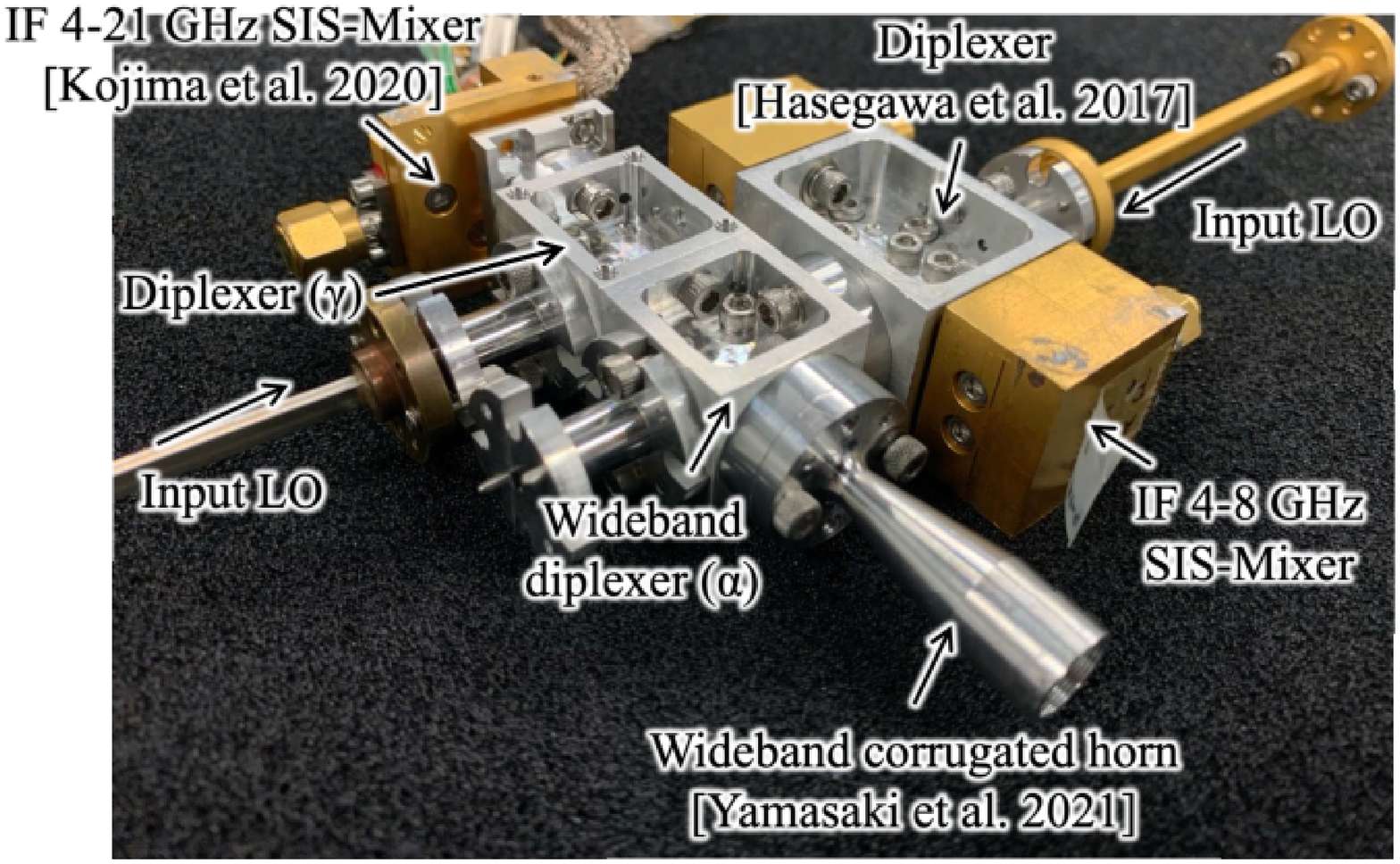}
 \end{center}
\caption{
A photograph of the receiver system assembled. The system consists of a wideband corrugated horn, the wideband multiplexer we developed here, a wideband SIS--mixer with 4--21\,GHz IF output, and a conventional SIS-mixer with 4--8\,GHz IF output. 
}\label{fig:Photograph_of_RF-circuits}
\end{figure}

\begin{figure*} [h]
 \begin{center}
  \includegraphics[width=16cm]{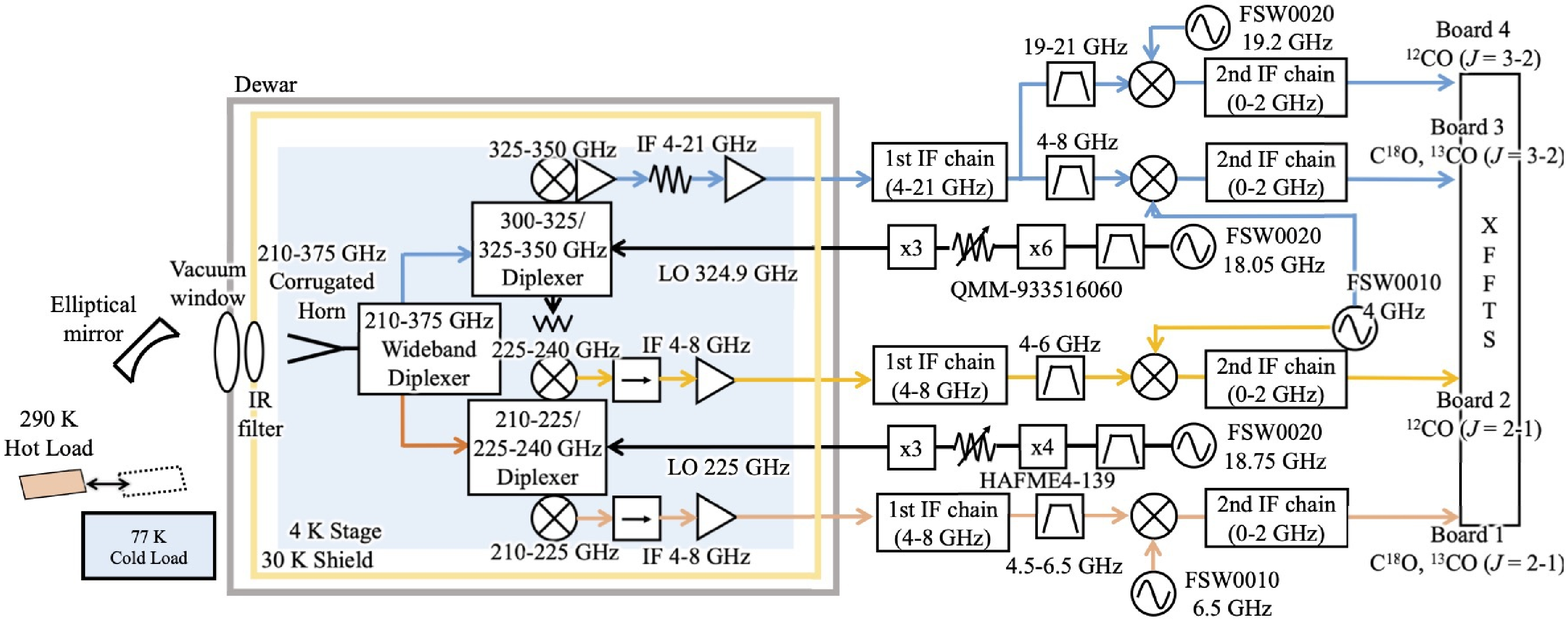}
 \end{center}
\caption{
The schematic diagram of the wideband receiver system for the observations. 
}\label{fig:Schematic_diagram_of_new_receive}
\end{figure*}

Due to the limited availability of some of the components, we modified some parts of the final configuration of figure\,\ref{fig:Schematic_Diagram} as shown below.
Using the developed multiplexer and the SIS--mixer developed by \citet{Kojima_wideband_RF_IF}, we developed a wideband receiver for simultaneous observations of CO isotopologue lines in the 230\,GHz and 345\,GHz bands.
Because the RF band of the developed IF 4--21\,GHz SIS--mixer supports the frequency range of 275--500\,GHz, it is used in the 345\,GHz band of our receiver.
In the 230\,GHz band, we used an SIS--mixer with IF 4--8\,GHz which employs a parallel-connected twin junction with a mirror symmetrical circuit about the bisection plane in the center of the waveguide (\cite{Asayama_sis_mixer}).
However, the multiplexer developed here is designed to use wideband IF even in the 230\,GHz band, so it cannot be used with a side-band--separated SIS--mixer in the 4--8\,GHz band.
Therefore, instead of the diplexer\,($\beta$), we decided to use another diplexer developed by \citet{Hasegawa_diplexer}, and change the LO frequency in the 230\,GHz band to 225\,GHz.
The corrugated horn, which couples the incoming 210--375\,GHz RF signal from the telescope to the waveguide input for the multiplexer is described in detail in another paper \citet{Yamasaki_PASJ}.
A photograph of the RF circuit for the simultaneous CO observations is shown in figure\,\ref{fig:Photograph_of_RF-circuits}, and the schematic diagram of the receiver for the observations is shown in figure\,\ref{fig:Schematic_diagram_of_new_receive}.
One of the output ports of the diplexer\,($\gamma$) in the 300--325\,GHz frequency band was terminated because there is no desired CO line.

The LO system is a Phase Matrix FSW 0020, and the active multiplier in the 230\,GHz band is a hxi HAFME 4 --139 $\times$4 multiplier and VDI WR--4.3 output tripler, while in the 345\,GHz band, a Quinstar QMM--933516060 $\times$6 multiplier and VDI WR--3.4 output tripler are used.
In our system, a BPF is inserted between the signal generator and the active multiplier chain to reduce the LO noise, and the details will be presented in another paper.
The LO frequencies were set to 225\,GHz and 324.9\,GHz for the frequency bands of the diplexers\,($\beta$) and ($\gamma$), respectively.
The 230\,GHz SIS--mixers' outputs are amplified at 4\,K by 4--8\,GHz cryogenic low noise amplifiers (CLNA), manufactured by Nitsuki, with cryogenic isolators, and feed into the 1st IF chains as shown in figure\,\ref{fig:Schematic_diagram_of_new_receive} at room temperature. The IF output is then downconverted by the 2nd mixers with the appropriate bandpass filters before each mixer, and the downconverted signals in the 0--2\,GHz range are transferred through the 2nd IF chains as shown in figure\,\ref{fig:Schematic_diagram_of_new_receive} to the spectrometers.
The 345\,GHz SIS--mixer's output is amplified at 4\,K by 3--22\,GHz CLNAs manufactured by Low Noise Factory.
A 10\,dB attenuator is inserted after the 1st CLNA instead of the wideband isolator covering the 3--22\,GHz which is not currently available.
The attenuator is able to avoid the saturation of the 2nd CLNA due to the wideband signal, and we install it at 4\,K to minimize the excess noise.
Then, the signal feeds into the 1st IF chain at room temperature.
A coaxial diplexer is attached to the room temperature part of the wideband IF 4--21\,GHz, and only the required bandwidth is extracted by BPFs.
The extracted signals are downconverted to 0--2\,GHz range by each mixer and also transferred through the 2nd IF chains in figure\,\ref{fig:Schematic_diagram_of_new_receive} to the spectrometers.
High-resolution wideband fast Fourier transform spectrometers (XFFTS \cite{XFFTS}) were used to obtain the spectra in the frequency range of 0--2\,GHz in the 2nd IF chain.
The resulting inputs are: board\,1: ${}^\text{13}$CO and C${}^\text{18}$O ({\sl J} = 2--1), board\,2: ${}^\text{12}$CO ({\sl J} = 2--1), board\,3: ${}^\text{13}$CO and C${}^\text{18}$O ({\sl J} = 3--2) and board\,4: ${}^\text{12}$CO ({\sl J} = 3--2).

\subsection{Laboratory measurement of the receiver noise temperatures with 4--21\,GHz IF output SIS-mixer}

\begin{figure} [h]
 \begin{center}
  \includegraphics[width=15cm]{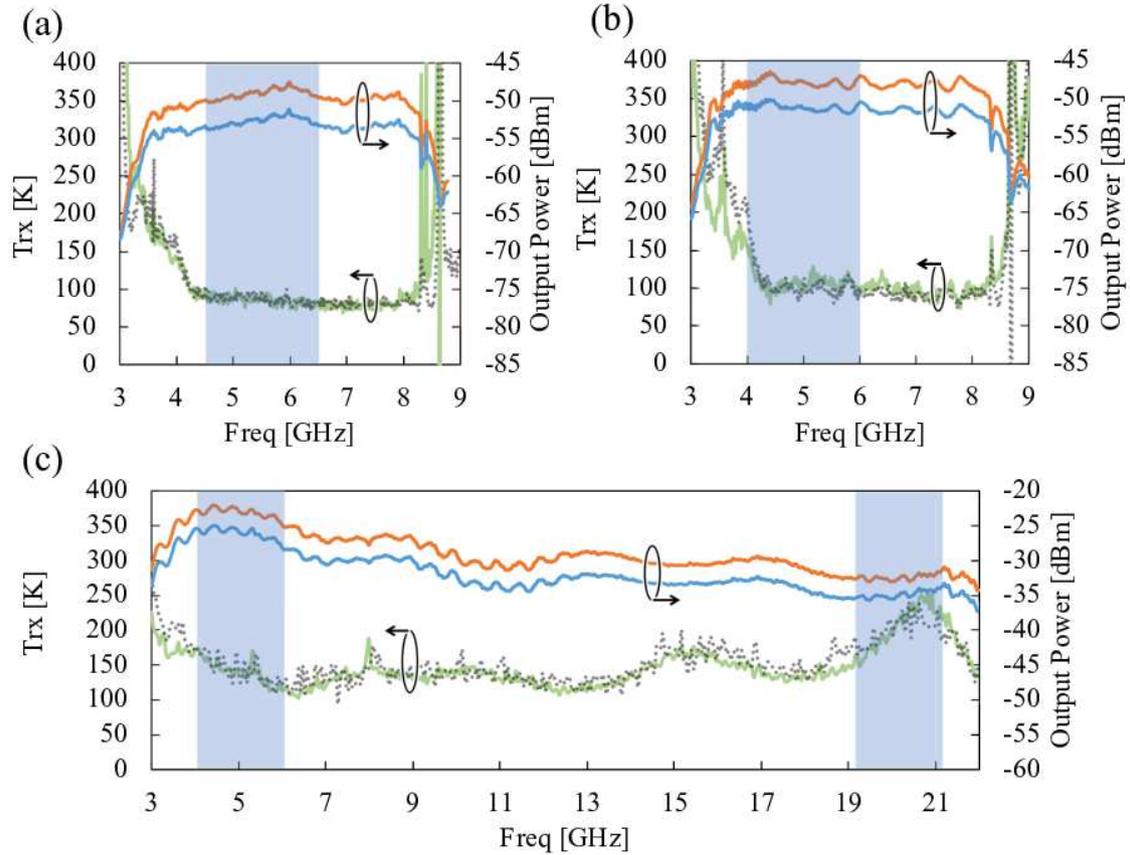} 
 \end{center}
\caption{
Measurement results of the SSB receiver noise temperature.  The horizontal axis shows the 1st IF frequency. Each graph shows the measurement results for (a) a lower side-band (LSB) with the LO frequency of 225\,GHz, (b) an upper side-band (USB) with the same LO frequency, and (c) an USB with the LO frequency 324.9\,GHz. The orange lines show IF output power when observing Hot load, the blue lines IF output power when observing Cold load, and the green line shows receiver noise temperature calculated from the Hot and Cold measurement.  The black dotted lines show the SSB equivalent noise temperatures by doubling the DSB temperature of each SIS--mixer measured separately. Blue square areas correspond to the spectrometer frequencies.
}\label{fig:Receiver_noise_temprature_SSB}
\end{figure}

Figure\,\ref{fig:Schematic_diagram_of_new_receive} shows a schematic diagram of the wideband receiver.
In the laboratory measurement, the single side-band (SSB) noise temperature was measured using a spectrum analyzer in the 1st IF section in order to check the characteristics of the entire IF as well as at the actual observed frequency.
The receiver noise temperature was measured using the standard Y--factor method with room temperature (T${}_\text{Hot}$ = 290\,K) and liquid nitrogen cooled (T${}_\text{Cold}$ = 77\,K) blackbody loads.
Each blackbody radiation was input using an elliptical mirror for laboratory measurements; a Keysight N9938A spectrum analyzer was used to check the IF output.
Figure\,\ref{fig:Receiver_noise_temprature_SSB} shows the results of the receiver noise temperature measurements in the laboratory.
The x-axis corresponds to the 1st IF frequency, the left side of the y-axis shows the calculated noise temperature value, and the right side of the y-axis shows the IF output power when Hot and Cold signals are input.
The noise temperature of the entire IF is about 70 to 100\,K in the 220\,GHz band, about 100\,K in the 230\,GHz band, and about 100 to 200\,K in the 330 to 345\,GHz band.
The blue square area corresponds to the spectrometer frequencies.
The noise temperatures of the double side-band (DSB) of individual SIS--mixers were also measured separately before the assembly.
The double of the DSB temperature, the SSB equivalent noise temperature, is shown by the black dotted line in figure\,\ref{fig:Receiver_noise_temprature_SSB}.
As estimated in section 2.6, we expect an extra noise corresponding to $\sim$1\,dB due to the insertion loss of the multiplexer.
However, the SSB equivalent noise temperature by the DSB measurement is quite similar to that of the multiplexer system.
One of the possible causes is that the SIS--mixers' performance tuning might be different between the measurements.
The independent measurement of the same SIS--mixers at NAOJ with the different measurement system resulted in the best DSB noise temperature of $\sim$45\,K within the same frequency range at the LO frequency of 325\,GHz, which is slightly better than the present DSB measurement (c.f., figure\,\ref{fig:Receiver_noise_temprature_SSB}(c)).
The optical misalignment at the present DSB measurement might be also a possible cause of the higher noise temperature.
Although the present measurements are not sensitive enough to estimate the insertion loss of the multiplexer, the overall performance of the system is reasonably high and seems to be consistent with the simulation result, at least not causing the significant increase of the noise temperatures of the system.

\newpage
\section{Commissioning observations using the developed receiver system on the 1.85--m telescope}
\begin{figure*} [h]
 \begin{center}
  \includegraphics[width=13cm]{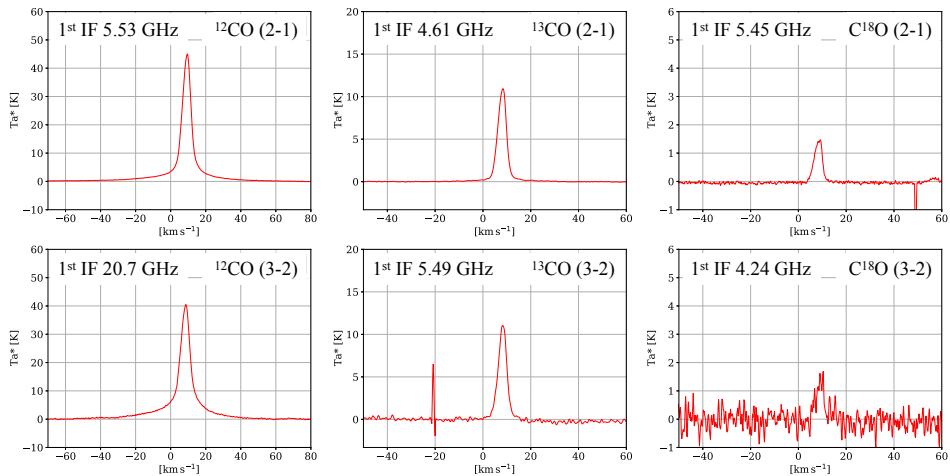} 
 \end{center}
\caption{
Observation results toward Orion\,KL. These CO lines with the transitions of {\sl J} = 2--1 and {\sl J} = 3--2 were observed simultaneously. The x-axis is the vlsr velocity, and the y-axis is the antenna temperature.
Spurious lines appear around 50\,km\,s{${}^\text{-1}$} of C${}^\text{18}$O ({\sl J} = 2--1) and --20\,km\,s{${}^\text{-1}$} of ${}^\text{13}$CO ({\sl J} = 3--2)
}\label{fig:OrionKL_band67_ps_20201114_freq}
\end{figure*}

\begin{figure*} [h]
 \begin{center}
  \includegraphics[width=12cm]{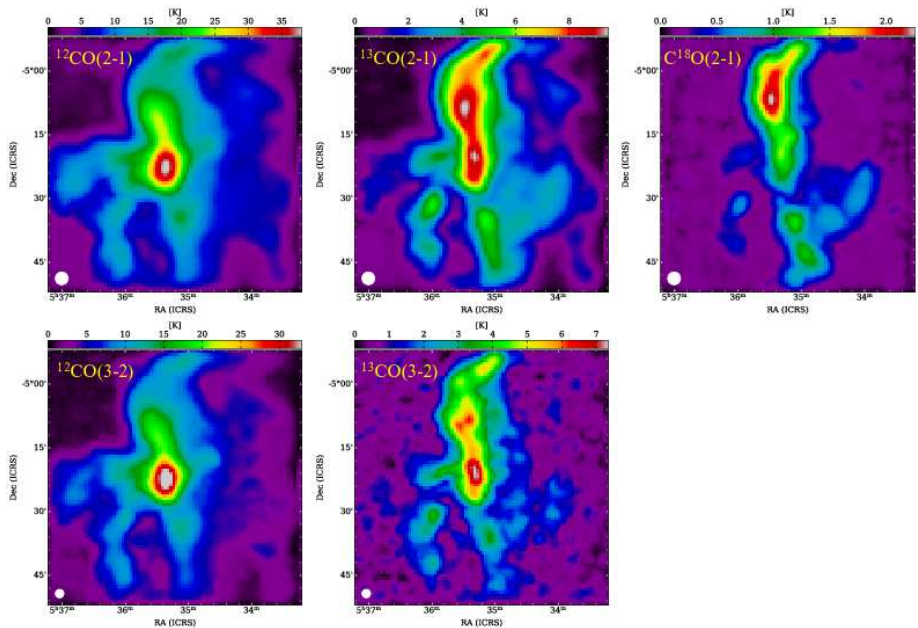} 
 \end{center}
\caption{
OTF maps of the peak brightness temperature toward Orion\,KL. The map size is $1^\text{°}\times1^\text{°}$. White circles show the beam size. Note that the C${}^\text{18}$O ({\sl J} = 3--2) lines are too faint to be mapped.
}\label{fig:OrionK_OTFmap}
\end{figure*}

\begin{figure*} [h]
 \begin{center}
  \includegraphics[width=13cm]{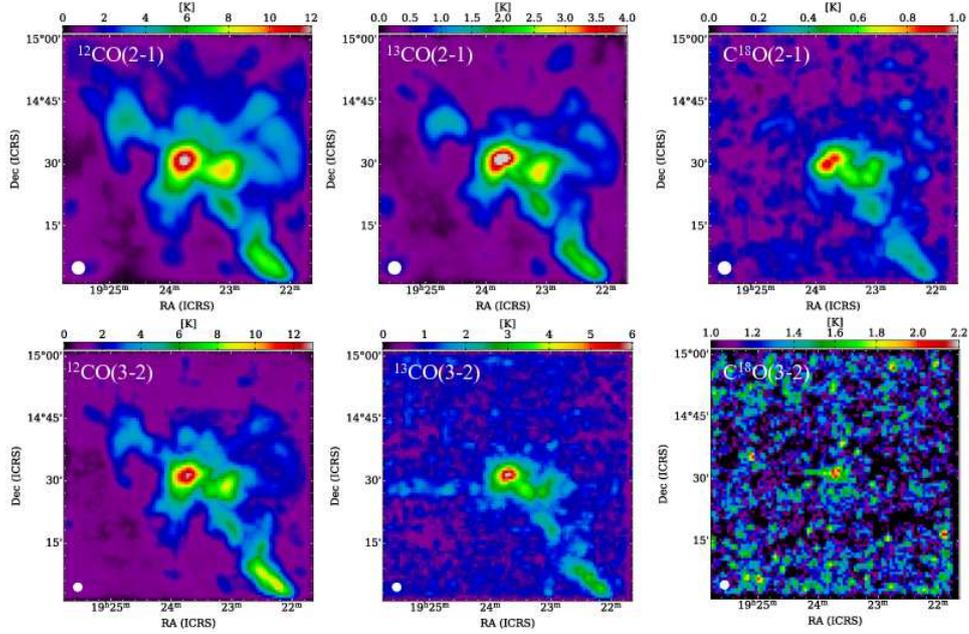}
 \end{center}
\caption{
OTF maps of the peak brightness temperature toward W\,51. The map size is $1^\text{°}\times1^\text{°}$. White circles show the beam size.
}\label{fig:W51_OTFmap}
\end{figure*}

The wideband receiver system is installed on the Osaka 1.85--m mm--submm telescope at NRO, which is developed and operated by OPU.
We also developed the antenna/receiver control system using distributed processing, which will allow us to flexibly respond to the introduction of new instruments.
The details of telescope control and antenna drive system can be found in \citet{Kondo_SPIE}.
We observed molecular clouds in Orion\,KL (RA\,=\,\timeform{05h35m14s.5}, Dec\,=\,--\timeform{05D22'30''} at J 2000.0) and succeeded in simultaneous observations of ${}^\text{12}$CO, ${}^\text{13}$CO, and C${}^\text{18}$O ({\sl J} = 2--1 and {\sl J} = 3--2).
Figure\,\ref{fig:OrionKL_band67_ps_20201114_freq} shows the chopper wheel calibrated spectra of six CO emission lines.
By comparing with the intensity of ${}^\text{12}$CO ({\sl J} = 2--1) in the 230\,GHz band by \citet{Nishimura_1p85m}, the main beam efficiency is considered to be comparable to the previous one.
The beam size and main beam efficiency are described in detail in Yamasaki et al. (2021).
The on-the-fly (OTF) mapping result toward the Orion\,KL region is shown in figure\,\ref{fig:OrionK_OTFmap}; although C${}^\text{18}$O ({\sl J} = 3--2) was too weak to be mapped, we succeeded in the simultaneous OTF observations of the other five emission lines.
The OTF mapping result toward the W\,51 region is also shown in figure\,\ref{fig:W51_OTFmap}.

\newpage

\section{Conclusion}
We have developed a waveguide-based wideband multiplexer covering a frequency range of 210--375\,GHz for the simultaneous observations of CO isotopologue molecular lines of {\sl J} = 2--1 and 3--2 transitions.
The results of this study can be summarized as follows:

\begin{enumerate}
  \item 
  We have designed and developed a wideband diplexer\,($\alpha$) with a fractional bandwidth of 56\,\%.
  The return loss is better than $\sim$17\,dB, and the cutoff frequency for dividing the RF signal is 275\,GHz.
  The basic design is a combined BLC and HPF diplexer proposed by \citet{Hasegawa_diplexer}, but to achieve a wider passband, sharper cutoff, and lower insertion loss, a compact cavity filter (\cite{Rosenberg_pseudo_HPF}) was used.

  \item 
  The diplexers\,($\beta$) and ($\gamma$) were developed to further divide the radio waves (210--275\,GHz and 275--375\,GHz bands, respectively) into two parts.
  They are also a combination of BLC and HPF and are designed to have a return loss of better than 20\,dB.

  \item 
  Three diplexers above were combined to form a multiplexer.
  The multiplexer can divide radio waves in the frequency band 210--375\,GHz into four bands: 215--235\,GHz (corresponding to C${}^\text{18}$O, ${}^\text{13}$CO and ${}^\text{12}$CO ({\sl J} = 2--1)), 245--265\,GHz, 305--321\,GHz, and 329--345\,GHz (C${}^\text{18}$O, ${}^\text{13}$CO, and ${}^\text{12}$CO ({\sl J} = 3--2)).
  We measured the {\sl S}--parameters of the multiplexer using a vector network analyzer and found that the return loss was better than 15\,dB in all frequency bands.
  The insertion loss was also consistent with the simulation results and was estimated to be approximately 1\,dB at\,4K.

  \item 
  The waveguide circuit was designed so that the output from the diplexers\,($\beta$) and ($\gamma$) and the LO signal are mixed and then input to the SIS--mixer.
  The output port from the diplexer\,($\gamma$) in the 329--345\,GHz band is connected to a SIS--mixer (Kojima et al. 2017) with an IF output bandwidth of 4--21\,GHz.
  This enables simultaneous detection of CO 3 emission lines of the {\sl J} = 3--2 transition. 
  The noise temperature of the entire IF is measured to be about 70 to 100\,K in the 220\,GHz band, about 100\,K in the 230\,GHz band, and 100 to 200\,K in the 330 to 345\,GHz band. 

  \item 
  Using the receiver systems on the 1.85m radio telescope, we succeeded in simultaneous observations of six CO isotopologue molecular ({\sl J} = 2--1, 3--2) lines toward Orion\,KL.
  This is also the first astronomical observation result with the SIS--mixer with an IF output bandwidth of 4--21\,GHz.
  The OTF mapping of $1^\text{°}\times1^\text{°}$ around Orion\,KL was also successfully done in five emission lines except for C${}^\text{18}$O ({\sl J} = 3--2), which is the weakest and noisiest emission line due to the atmosphere.
  In addition, the OTF mapping toward W\,51 with the same mapping size was successfully done in six emission lines.
\end{enumerate}

In the ALMA Development Roadmap (\cite{ALMA_development_roadmap}), which describes a roadmap for future developments of ALMA, the wideband RF and IF are listed as one of the important development items.
Our study pursues wideband heterodyne reception in mm and submm wavelength, and the actual observation on a telescope has been achieved.
We believe that this research result is a fundamental achievement for future wideband observations by high-performance radio telescopes such as ALMA.

%\section{Acknowledgement}

%\newpage

\begin{ack}
The authors would like to thank Issei Watanabe and Satoshi Ochiai of NICT.
They helped us to measure the {\sl S}--parameter of waveguide circuits.
The authors would like to also thank Shigeru Fuji of NRO and Toshihisa Tsutsumi of Yamaguchi University for the kind help at Nobeyama.
The authors are very grateful to the entire staff of the NRO and everyone related to the 1.85m telescope.
We are grateful that Shohei Ezaki of NAOJ and Masayuki Ishino of KMCO helped us to develop the wideband diplexer ($\alpha$).
We thank the referee, Dr. Noriyuki Kawaguchi, for helpful comments.
This work was supported by JSPS KAKENHI Grant Numbers JP15K05025, JP26247026, JP18H05440, and JP20J23670.
This work was also supported the grant of Joint Development Research supported by the Research Coordination Committee, National Astronomical Observatory of Japan (NAOJ), National Institutes of Natural Sciences(NINS).
\end{ack}

\appendix 
\section*{The final design of diplexers}
\begin{figure*} [p]
 \begin{center}
  \includegraphics[width=16cm]{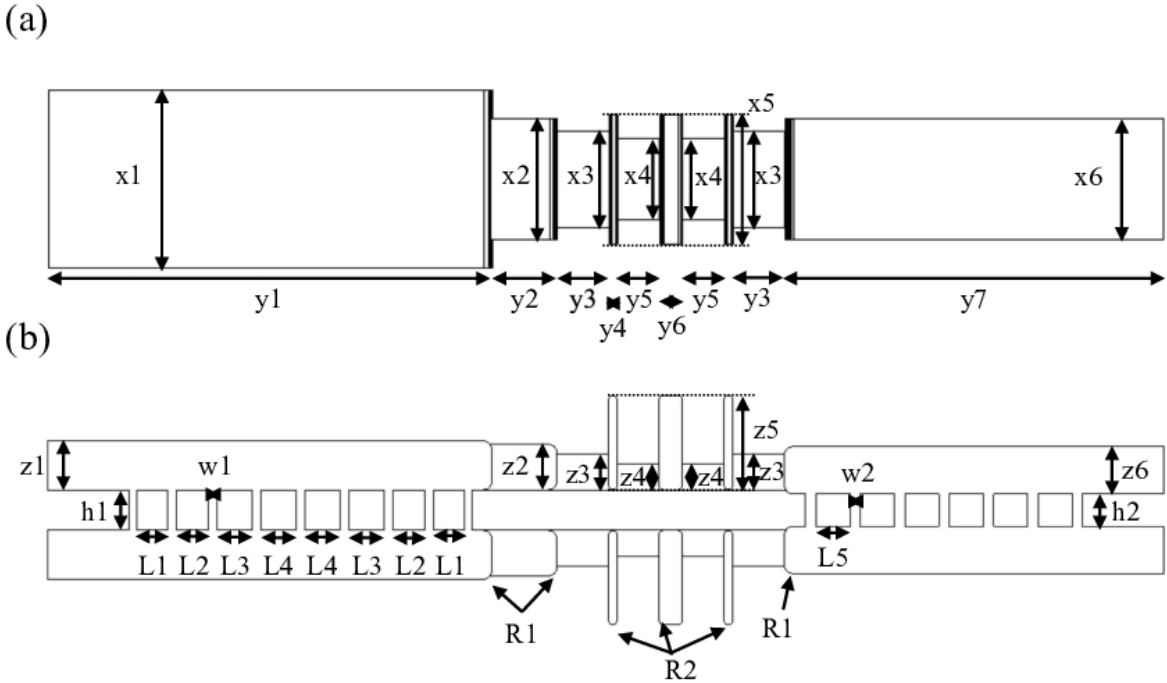}
 \end{center}
\caption{
Dimensions of the wideband diplexer\,($\alpha$) when the two split--blocks were assembled.
(a) The top view.
(b) The cross-sectional view from the side. The parameters starting with "R" stand for the corner radius.
This design uses different BLCs before (left) and after (right) the HPF.
The design of the above BLCs are described in detail in \citet{Masui_diplexer}.
}\label{fig:the_final_design_of_diplexer_a}
\end{figure*}

\begin{figure*} [p]
 \begin{center}
  \includegraphics[width=15cm]{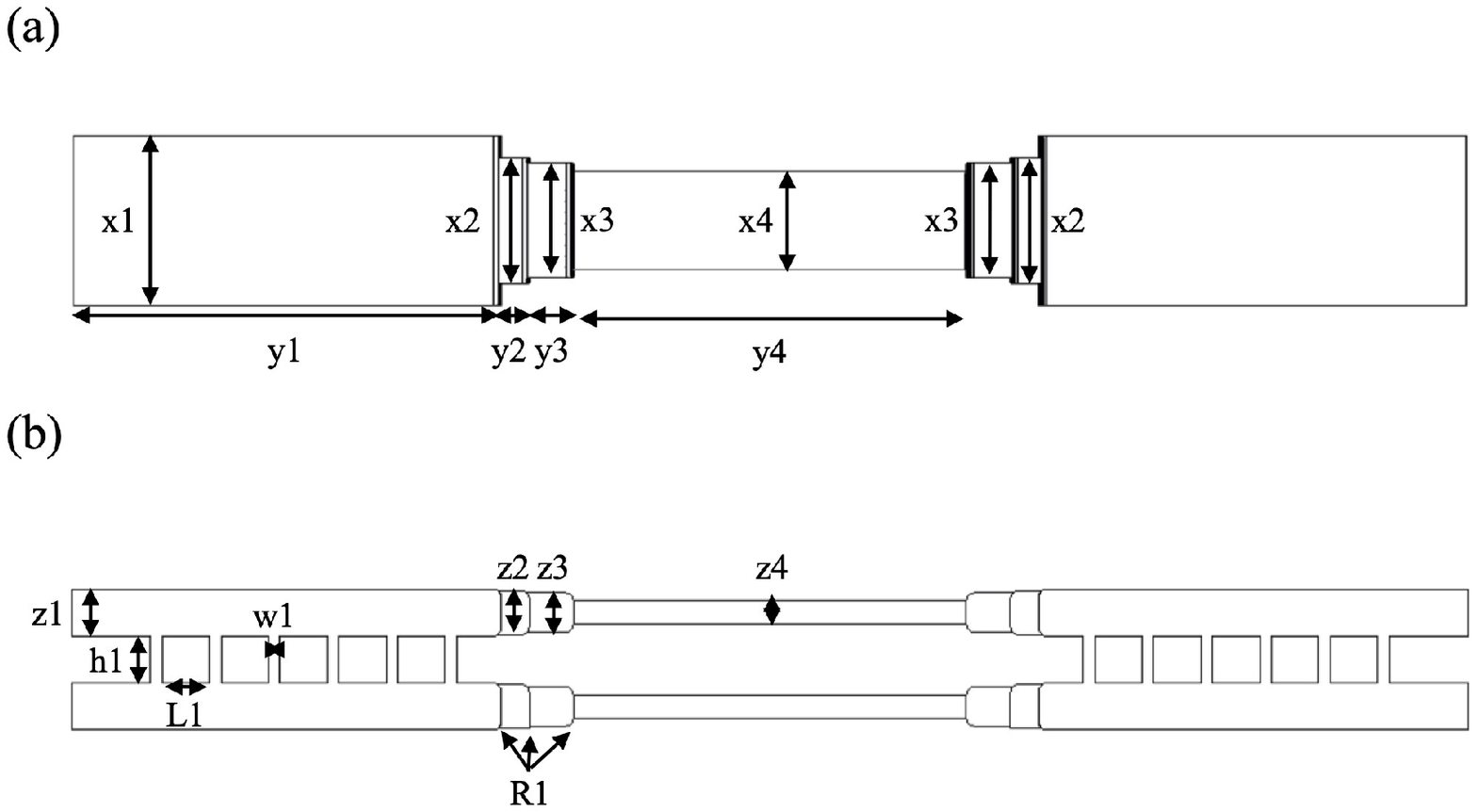}
 \end{center}
\caption{
Dimensions of the diplexer\,($\beta$) when the two split--blocks were assembled.
(a) The top view.
(b) The cross-sectional view from the side. The parameters starting with "R" stand for the corner radius.
The diplexer\,($\beta$) was designed to be symmetric.
}\label{fig:the_final_design_of_diplexer_b}
\end{figure*}

\begin{figure*} [h]
 \begin{center}
  \includegraphics[width=15cm]{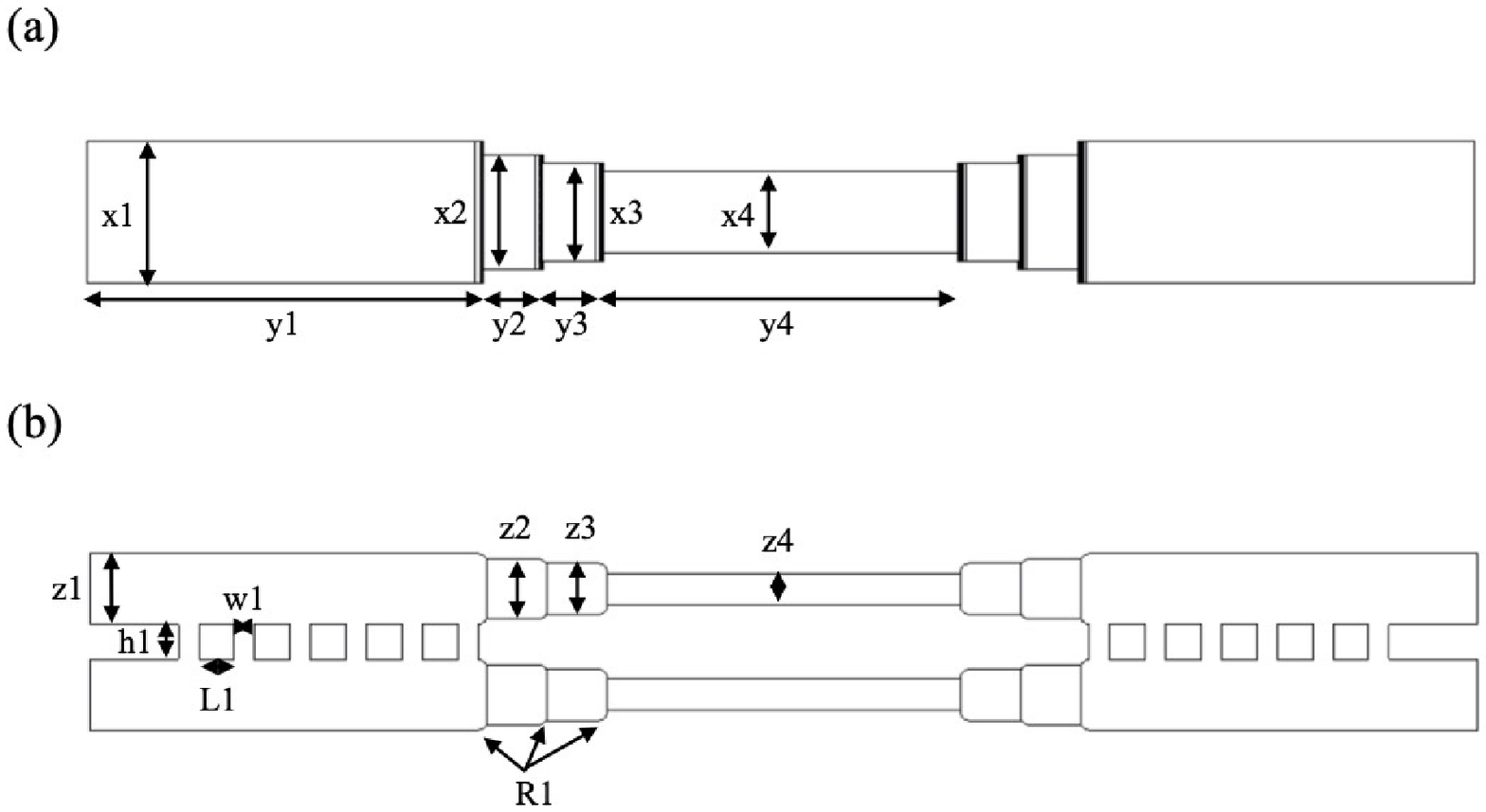}
 \end{center}
\caption{
Dimensions of the diplexer\,($\gamma$) when the two split--blocks were assembled.
(a) The top view.
(b) The cross-sectional view from the side. The parameters starting with "R" stand for the corner radius.
The diplexer\,($\gamma$) was designed to be symmetric.
}\label{fig:the_final_design_of_diplexer_c}
\end{figure*}

In this Appendix, the structural parameters of the developed wideband diplexer\,($\alpha$), ($\beta$), and ($\gamma$) are shown in table\,\ref{tab:the parameters of diplexers} and in figures\,\ref{fig:the_final_design_of_diplexer_a}, \ref{fig:the_final_design_of_diplexer_b}, and \ref{fig:the_final_design_of_diplexer_c}.
The design of the wideband diplexer\,($\alpha$) uses different BLCs before and after the HPF.
The designs of the above BLCs is described in detail in \citet{Masui_diplexer}.

\begin{table}[p]
%\begin{minipage}
  \tbl{The size parameters of the diplexers.}{%
  \begin{tabular}{llll}
      \hline
       Parameter & figure\,\ref{fig:the_final_design_of_diplexer_a} [mm] & figure\,\ref{fig:the_final_design_of_diplexer_b} [mm] & figure\,\ref{fig:the_final_design_of_diplexer_c} [mm]\\ 
      \hline
       \multirow{1}{*}{x1}& 1.092 & 1.092 & 0.8\\
       \multirow{1}{*}{x2}& 0.74 & 0.80 & 0.065\\ 
       \multirow{1}{*}{x3}& 0.60 & 0.73 & 0.55\\
       \multirow{1}{*}{x4}& 0.50 & 0.63 & 0.466\\
       \multirow{1}{*}{x5}& 0.80 & NA & NA\\
       \multirow{1}{*}{x6}& 0.74 & NA & NA\\
       \multirow{1}{*}{y1}& 2.73 & 2.73 & 2.246\\
       \multirow{1}{*}{y2}& 0.40 & 0.18 & 0.34\\
       \multirow{1}{*}{y3}& 0.32 & 0.28 & 0.34\\
       \multirow{1}{*}{y4}& 0.05 & 2.5 & 2.0\\
       \multirow{1}{*}{y5}& 0.26 & NA & NA\\
       \multirow{1}{*}{y6}& 0.14 & NA & NA\\
       \multirow{1}{*}{y7}& 2.33 & NA & NA\\
       \multirow{1}{*}{z1}& 0.30 & 0.30 & 0.4\\
       \multirow{1}{*}{z2}& 0.28 & 0.28 & 0.34\\
       \multirow{1}{*}{z3}& 0.22 & 0.25 & 0.29\\
       \multirow{1}{*}{z4}& 0.16 & 0.15 & 0.18\\
       \multirow{1}{*}{z5}& 0.58 & NA & NA\\
       \multirow{1}{*}{h1}& 0.25 & 0.30 & 0.20\\
       \multirow{1}{*}{h2}& 0.20 & NA & NA\\
       \multirow{1}{*}{w1}& 0.05 & 0.075 & 0.116\\
       \multirow{1}{*}{w2}& 0.063 & NA & NA\\
       \multirow{1}{*}{L1}& 0.19 & 0.30 & 0.2\\
       \multirow{1}{*}{L2}& 0.20 & NA & NA\\
       \multirow{1}{*}{L3}& 0.22 & NA & NA\\
       \multirow{1}{*}{L4}& 0.22 & NA & NA\\
       \multirow{1}{*}{R1}& 0.025 & 0.05 & 0.05\\
       \multirow{1}{*}{R2}& 0.05 & NA & NA\\
      \hline
    \end{tabular}}\label{tab:the parameters of diplexers}
\begin{tabnote}
%The target value in the design of the characteristic required for each circuit is shown.
\end{tabnote}
%\end{minipage}
\end{table}

\newpage

\newpage

%%%
% See the manual for the detail.
%%%

\end{document}